# New Physical Mechanism for Lightning

**Sergey N. Artekha • Andrey V. Belyan**

*Space Research Institute of RAS, Moscow, Russia*

Corresponding author: Sergey N. Artekha
e-mail: Sergey.Arteha@gmail.com
ORCID: 0000-0002-4714-4277

**Abstract** The article is devoted to electromagnetic phenomena in the atmosphere. The set of experimental data on the thunderstorm activity is analyzed. It helps to identify a possible physical mechanism of lightning flashes. This mechanism can involve the formation of metallic bonds in thunderclouds. The analysis of the problem is performed at a microphysical level within the framework of quantum mechanics. The mechanism of appearance of metallic conductivity includes the resonant tunneling of electrons along resonance-percolation trajectories. Such bonds allow the charges from the vast cloud charged subsystems concentrate quickly in lightning channel. The formation of metal bonds in the thunderstorm cloudiness is described as the second-order phase transition. A successive mechanism for the process of formation and development of the lightning channel is suggested. This mechanism is associated with the change in the orientation of crystals in growing electric field. Possible consequences of the quantum-mechanical mechanism under discussion are compared with the results of observations.

# 1 Introduction

Electromagnetic phenomena are presented in many processes in the world around us. A striking example here is the electrical activity, which manifests itself in crisis atmospheric phenomena. Among them storms are the most widespread. Lightning discharges accompany cyclones, including tropical (hurricanes, typhoons), tornadoes, cumulonimbus clouds (including those containing hail). This article is devoted to the study of processes occurring in thunderclouds.

According to Wilson's hypothesis [1], storms represent generators of some global electric circuit [2]. Universal diurnal variation of gradient of the potential is observed across the Earth, the so-called "Carnegie curve" [3]. It has a maximum at 19-00 GMT, that can be connected with the activity of thunderstorms in the global electrical circuit. Note the following: for the time of this maximum, the sun is located at the highest point (high ionization) over the meridian with the lowest value of the geomagnetic field. Note another observation. The portion of intracloud lightnings increases (from 0.5 to 0.9) at a displacement to the equator, compared with a portion of cloud-to-ground discharges. This can also be connected with the Earth's magnetic field, which is parallel to the earth's surface near the equator (even statistically, motions of charged particles along and across the field are non-equivalent).

One of the important physical problems in crisis atmospheric processes is to determine mechanisms for reproduction of charged particles. The most famous mechanisms include ionization by radioactive substances near the earth's surface, which is effective at altitudes of up to three kilometers, ionization by cosmic rays, ionization of aerosols by hard ultraviolet, etc. In a number of phenomena there is an explicit mechanism for the formation of ions: for example, over powerful fires or volcanoes. In many vortex phenomena, one can see the mechanism of reproduction of charged particles by friction or collision of aerosols (for example, for dust storms, snow whirlwinds and tornadoes, sometimes formed between clouds). Note that lightnings are also observed during volcanic eruptions and dust storms. Perhaps, some correlation with a similar mechanism, plus the presence of aerosols and more intense upward thermal currents cause relatively high intensity of thunderstorms above mainland Earth (10 times compared with the Ocean). There exists some asymmetry between anions and cations leading to possible mechanisms of charge reproduction (attaching of ions to hydrometeors and electrification of hydrometeors in the process of phase and microstructure transformations).

Although the study of thunderstorms is began by B. Franklin and M.V. Lomonosov, at present the work is far from its completion. Many hypotheses were suggested about the reasons for the separation of electric charges in thunderclouds, but none of them individually covers all observational data (or there are contradictions). We mention some of the most well-known hypotheses on the basic mechanisms of separation of electric charges (for more details, see [4]). These include the following schemes: Elster's, Geitel's [5] and Wilson's [6] schemes that in the gravitational field larger particles of precipitation with one sign are separated from smaller particles in clouds with another sign; Grenet's and Vonnegut's scheme about the induction mechanism - light ions with an ascending stream settle on the particles and create a positively charged layer, and a negatively charged layer is formed on the boundary [7-9]. The separation of charges can occur when charged particles are emitted in the process of freezing, at breaking up droplets and spalling pieces from hoarfrost [10]. The convective charging mechanism is based on the vertical transport of charges generated by the corona due to the irregularities of the electric field above the earth's surface. The article [11] presents the quantitative theory of generation and of charge separation when supercooled water droplets in clouds contacted with the underside of ice pieces, polarized in the initial electric field of the Earth. The positive charge is rise upward clouds with the lifting force, while hailstones carry a negative charge to the lower part of the cloud. One more theory of thundercloud electrification is presented in [12]. The polarization of ice crystals exhibiting ferroelectric properties leads to the generation of an electric field and to the appearance of screening charges on the upper and lower boundaries of the cloud. These charges, being freed when the physical properties of cloud particles change, can participate in the formation of lightnings.

A large number of papers have been devoted to the physics of the lightning discharge (see monographs [13, 14]). However, not all issues can be considered to be closed here. After a very long process of generation and separation of charges, the following problem is the most difficult to explain lightning flashes. How, despite the low static conductivity of the intra-cloud dielectric medium, charges of one and the same sign are concentrating into a narrow lightning channel on average during thousandths of a second (the cases from $10^{-5}$s to $10^{-1}$s is recorded)? These charges synchronously flow down from isolated particles of a huge cloud space, which occupies several cubic kilometers! Fractal dynamics of intracloud microdischarges and formation of the drainage system of the electric charge transfer are investigated in [15-17].

Another difficulty in explaining the processes of lightning discharges is associated with the absence of electrodes and with low breakdown fields. Despite the increased interest in such studies, this problem cannot yet be considered to be closed. The contribution of the influence of metastable oxygen and nitrogen molecules to the lowering of the threshold for the electrical breakdown of the cloud medium was taken into account in [18] and it was concluded that the contribution of these metastable states is unlikely to be significant for the initiation of lightning streamers. In [19] the simulation of the physical process of preliminary lightning breakdown was carried out in simplified geometry with allowance for the change in resistance due to heating of the lightning channel and corona discharges.

The storm cloud as a whole is an active strongly nonequilibrium system, in which there is no noticeable long-range order (although it would be worth mentioning the forces that keep the cloud structure together, often in spite of other forces, for example, in charged layers). We will be interested in the dynamic manifestations of short-range order. Recall that this is the presence of some regularity or correlation on a certain scale $R_c$ for some characteristic (distance, composition, orientation, polarization, etc.) among neighboring atoms or molecules. For example, liquids can possess the short-range order, (despite the chaotic and much faster movement of molecules compared with turbulent motion).

This article is devoted to the problem of initiation and development of the lightning discharge. The main emphasis is placed on the inclusion of microphysical processes and small-scale phenomena in the description, taking into account the collective behavior of particles and the influence of this behavior on the system as a whole. The main goals of this paper are as follows:

- To attract the attention of researchers on some observations concerning the lightning phenomena;

- To discuss a possible mechanism that allows charges to flow down from the vast cloud of charged subsystems for lightning flashes;

- To suggest a description of the formation of metallic bonds in thunderclouds as a second-order phase transition;

- To suggest a mechanism for formation of the lightning channel.

We will try from a single position (namely, from the viewpoint of our hypothesis) to offer the explanation of a number of observable phenomena, mentioning how to describe them, modern representations or existing hypotheses where it is required for objectivity.

## 2  Metal Bonds. Phase Transition

A necessary prerequisite for the occurrence of thunderstorms is the accumulation of a sufficient number of charges in a certain area of the atmosphere. Since the ion formation is strongly correlated with the solar radiation, this is a possible cause that the maximum of thunderstorms is observed in the summer during any year and in the afternoon - during any day globally on the Earth (a minimum - just before sunrise). Ions act as condensation nuclei, where latent heat of vaporization is released. This is the trigger mechanism for conversion of enormous reserves of the thermal energy (steam) into the kinetic energy. This in turn, together with the convection caused by direct solar heating of the earth surface, promotes the subsequent separation of charges.

Although it is not obvious, the role of electromagnetic forces is noticeable in many atmospheric phenomena. Evaluation of the density of electric forces and motions under the influence of electromagnetic forces leads to suggestion of some ordering of charged particles and their motions in clouds: some analogy of a dynamic crystal in statistical equilibrium [20]. Electromagnetic forces are involved in the maintenance of a quasi-equilibrium structure of cloud layers and macroscopic movements [21, 22]. But explicitly they manifest themselves during lightning flashes.

Lightnings are substantially different from artificial laboratory discharges between electrodes. For example, lightnings often occur with the field strength, much lower than breakdown values [23]. The process has a stepped nature, and the characteristics of pulses of HF-radiation indicate that the radiation occurs on sizes of the order of 10-100 meters. There exists also a certain minimum length of lightning discharges - the order of several hundreds of meters.

At present, one of the most advanced theories describing ionization of the atmosphere for the implementation of lightning is the theory of breakdown on runaway electrons initiated by cosmic rays (or arising extensive air showers - EAS - electron avalanches). That model is described in detail in [24-26]. Undoubtedly, wide atmospheric showers and breakdown on runaway electrons can play an important role in the ionization process and in the initiation of discharges. Undoubtedly, extensive atmospheric showers and breakdown on runaway electrons can play an important role in the ionization process and in the initiation of discharges. However, as it seems to us, this mechanism is not the only and exhaustive, since it does not cover the whole set of experimental data on the behavior of lightning. We list some points.

It is well known that the lightning discharge is by no means rectilinear, determined by the field gradient (eg, vertical). Lightning uses a single channel, but does not propagate by multiple parallel courses (especially obvious that this is not uniform discharge or not glow of the whole cloud). That is why the attraction of extensive air showers and runaway electrons as the sole mechanism is very problematic for explanation of lightnings. Cosmic particles move by parallel courses towards the Earth. Extensive atmospheric showers (mainly electrons) occupy a width of 100-1000 meters or more. However for some reason, the entire cloud (often a multi-layered charged structure) does not discharge evenly (does not neutralized), but the discharge is concentrated in a narrow lightning channel (only a very small number of particles from the EAS could get into its small volume). Further, in nature there are observed not only cloud-to-ground lightnings and cloud-to-ionosphere lightnings (vertical - 25%), but also cloud-to-cloud lightnings (conditionally horizontal - 75%, including between multilayer structures), that is, discharges occur in a wide range of directions of the electric field. However the electrons could accelerate to the state of "runaway electrons" only along the field, but not perpendicular to it, or not against the field (where the "initiating" fast electrons decelerate). In this case, the breakdown voltages would coincide with the laboratory values, since cosmic rays are also observed in the laboratory. Equipment for discharge monitoring does not belong to the high-tech equipment and is available in many institutes and universities of the world, located at very different altitudes above sea level. However, no one has yet discovered any significant dependence of the breakdown voltage on the altitude. The statement, that the minimal lightning length for observation of this effect should be larger than the critical size (artificially taken from the exponent) $L \sim 60$ meter, is not convincing enough. The breakdown voltage is determined per unit length and the theory does not contain a threshold dependence (the greater the path traveled, the greater the energy of the secondary particles). After all, the result should depend directly on the flux density and the energy of the initiating particles. In the end, according to theory, the mechanism reduces to the degree of ionization (and this is an experimentally measurable quantity). Therefore, it would be worthwhile to compare the directions of the fields, the directions of the fluxes, the density and energy of the EAS particles at the inlet and outlet (in fact, the experimental correlation is not 100%). All the above difficulties are most noticeable in explaining intracloud lightnings. In a number of photographs, the structure of lightning visually resembles a statistically self-similar (up to 2-3 orders) root system. Despite the long-range character of the Coulomb field, often the movement of lightnings (chaotic enough in directions) includes reverse trajectories (and even the similarity of loops), that is, it occurs at some path sections against the mean field. All abovementioned rather indicates that lightning represents some collective process formed in the unified manner, in which new properties of a cloudy atmosphere in the future lightning channel just play a key role. Indirectly, this is evidenced by a noticeable increase in the brightness of clouds during lightning: an increase in reflectivity can be associated with possible metallization. If this had been due solely to the ordering of ice crystals in an electric field, the increase in brightness would have been significantly anisotropic. Hence, this phenomenon is explained by an additional change in properties (including those reflecting) of storm clouds themselves. From the viewpoint of the electrical breakdown (impact ionization), the "Catatumbo Lightnings" are also incomprehensible (up to 200 days a year over Lake Maracaibo), which are not accompanied by thunder at all. The volt-ampere characteristic of a lightning discharge (with a "break in the curve") and a broad, relatively smooth spectrum of radiation energy, evidence indirectly in favor of a phase transition. And X-ray flashes during lightning discharges (motion between neighboring regions with metallic properties) are analogous to those in a conventional X-ray tube, when the electrons accelerated between the electrodes hit the metal surface of the anode. Even an X-ray flash occurs just before the registration of the electric pulse during the passage of the next conducting section by the lightning (the process can be described with the help of a mini return strike). The collision of ions with a collectivized (metal bond) region can be one of the reasons for the generation of neutrons. The following fact testifies in favor of the formation of the metal bond. Despite the uneven propagation, the entire channel restores the glow during the movement of the leader, including the previously passed sections, that is, the unified metallized channel is involved into an electrical circuit (which leads to equalization of its potential). Thunderstorms are identified by a number of criteria, both from satellites and meteorological radars, for example, they have the highest radar reflectivity (therefore, not only the sizes of ice sheets and droplets are important), that also indirectly indicates the presence of free charges and a single metal-like structure.

Considering the foregoing, lightning resembles a short-term light flash arising from the action of an electric current as a result of a statistically appearing conductive metal channel for the Catatumbo Lightning or the successively propagated glow of several conducting metal channels with electrical breakdowns between them for ordinary lightning (in this case the breakdown of an air is accompanied by pressure waves - the thunder, that can be explained with the streamer theory). It is not enough just to create the required electric field strength for an adequate laboratory simulation of lightning. First of all, lightning is associated with static electricity. It is necessary to take into account the actual microphysical structure of the cloud layer, including not only the quantity, density, shape and size of the crystals, ion composition, but, most importantly, the required density of excess electric charge.

Note that not only purely metallic crystals (perfectly ordered and unlimited in space) can conduct electricity, but also limited polycrystalline conductors (having obviously anisotropic structure), and even disordered alloys and liquid mercury (not only ordering absent here, but also the fixed structure). It is possible that for the manifestation of this property a certain short-range order is sufficient and the concept of "resonance-percolation trajectories, i.e. tunneling paths along which there is no attenuation" is applicable [27, 28]. In this case, the average transparency coefficient is formed from unlikely configurations corresponding to almost complete transparency of the trajectory (although the metallic conductivity of the droplet-crystal cloud seems even more exotic than the good conductivity of liquid mercury).

The appearance of a metal bond in the charged subsystems of thunderstorm clouds can be represented as a second-order dielectric-metal phase transition [29], which will be described below. (However, attributes of a first-order phase transition can also be observed, for example, the growth of sizes and charges of cloud particles before the first flash of lightning; see similar examples in [30]). Such chaotic metallic bonds allow charges to flow very quickly from enormous space occupied by the charged cloud subsystem to the arising lightning channel, the formation of which will also be explained later.

From the viewpoint of the classical nonequilibrium thermodynamics, to describe the transition we can go from the Helmholtz free energy $F_1$ and entropy $S$ (additive quantities) to their densities and can take their values depending on the coordinates and time for all local variables: $F_1/N = F(\mathbf{r},t) \equiv F$, $S/N = s(\mathbf{r},t) \equiv s$, $N$ - is the total number of particles, $P(\mathbf{r},t) \equiv P$ - is the local pressure, $T(\mathbf{r},t) \equiv T$ - is the local temperature. Substituting the concentration of resonant structures $n(\mathbf{r},t) \equiv n$ instead of volume $V$: $V \to N/n$, we obtain:

$$dF = \frac{P}{n^2}dn - sdT .$$

As is known, the expansion of $F$ near the point of transition in the order parameter $\eta$ has the form:

$$F(T,n,\eta) = F_0(T,n) + A(T,n)\eta^2 + B(T,n)\eta^4 ,$$

where $A(T,n) = a(T)(n-n_c)$, $B(T,n) \approx B(T)$, $n_c$ - is some critical concentration. As a result

$$F(T,n,\eta) = F_0(T,n) + a(T)(n-n_c)\eta^2 + B(T)\eta^4 .$$

In principle, the effect of an external inhomogeneous alternating field **E** (or **H** from the arising current) can be taken into account by introducing of the term $-\hat{\eta}\dfrac{E}{n}$ (or $-\hat{\eta}\dfrac{H}{n}$). As the order parameter, it is necessary to choose a quantity that varies significantly when passing through the critical point. For example, we can choose the value of $|\psi(\mathbf{r},t)|^2$, where $\psi(\mathbf{r},t) \equiv \psi$ - is the wave function of a fraction of collectivized electrons (ideally, taken from the time-dependent Schrödinger equation). (Or we can choose the transparency coefficient for barriers involved in short-range order - it also varies from practically zero to almost one.) We use the fact that in a thundercloud excess electrons are at a large distance from each other (near to not the same, but different atoms). Then in the first approximation [31], this value can be presented as a product of potentially possible single-particle wave functions $\psi = \prod_{i=1}^{N}\psi_i$ of electrons (conditionally free and included in ions and in the hydrated formations) that can be involved in the creation of the metallic conductivity of a resonance region. Of course, we can take other, more accurate approximations for $\psi$ (antisymmetrized product of real single-particle wave functions). Integrating over the region $V$, we obtain in the case of localized electrons: $\eta = 0$. If the short-range order is formed with collectivized electrons in some area of $v$, then the integration over this area of the short-range order gives $\eta = 1$. In the intermediate case, the integration over a larger area gives $0 < \eta < 1$ (depending on the region of overlap between such resonant regions).

Similar to the description of the superconducting transition, we have:

$$F_1 = F_0 + \int\left(\dfrac{\hbar^2}{4m}|\nabla\psi|^2 + a|\psi|^2 + \dfrac{b}{2}|\psi|^4\right)dV,$$

the equilibrium value is $|\psi|^2 = \dfrac{\alpha}{b}(n - n_c)$ with $\alpha = \dfrac{a(T)}{v}$, $b = \dfrac{B(T)}{v}$. The free energy change is:

$$F_1 - F_0 = -V\dfrac{\alpha^2}{2b}(n - n_c)^2.$$

Taking into account the lightning current ($\mathbf{j} = \sigma\mathbf{E}$) and arising field, the final expression is:

$$F_1 = F_0 + \int \left( \frac{B^2}{8\pi} + \frac{\hbar^2}{4m} \left| (\nabla - \frac{2ie}{\hbar c} \mathbf{A})\psi \right|^2 + a|\psi|^2 + \frac{b}{2}|\psi|^4 \right) dV$$

$$\text{rot } \mathbf{B} = \frac{4\pi}{c}\mathbf{j}, \quad \mathbf{j} = \frac{-ie\hbar}{2m}(\psi^*\nabla\psi - \psi\nabla\psi^*) - \frac{2e^2}{mc}|\psi|^2 \mathbf{A}.$$

The equation for the transition line is:

$$\frac{1}{4m}(-i\hbar\nabla - \frac{2e}{c}\mathbf{A})^2\psi + a\psi + b|\psi|^2\psi = 0.$$

Of course, the Landau theory is used as the first approximation only (for more correct calculation, it is necessary to take into account long-range correlations).

In general, the study of metal-insulator transitions drastically changed the traditional separation into metals and dielectrics by the type of electron spectrum and by the filling of zones of collectivized electrons [30]. For example, the ions $V^{3+}$ and $V^{4+}$ of Magneli phases $V_2^{3+}V_{k-2}^{4+}O_{2k-1}$ are arranged in the metal phase randomly (!), while in the insulating phase - orderly. Note that gaseous metals can also exhibit metallic properties [32]. Thus, nature experimentally confirms the possibility we are suggesting.

## 3 A Possible Mechanism of Metal Conductivity

Since we assume that in the phenomenon under consideration a change in the state and properties of the system as a whole occurs, so we will use the quantum-mechanical approach (instead of the usual kinetic description). It is known that the velocity of the molecules themselves in a solid, liquid or gas is hundreds of meters per second, but this does not interfere with the applicability of the methods of quantum mechanics. Observing lightnings (non-rectilinear) in thunderclouds, a significant local anisotropy of the electrical conductivity of clouds can be supposed. This supports the model of metal phase with quasi-one-dimensional conduction band. At once we emphasize that talking is not at all one-dimensionality of the system, since there are no phase transformations in the one-dimensional case. Even when considering a thunderstorm cloud system as a whole, we have a complex stochastic system of conducting channels (fractal), the dimension of which is obviously greater than one. On the scales of the local conducting section, the system is three-dimensional; otherwise there would be no collection of charges from the volume of the cloud. However, the current tries to flow along the path of least resistance, and therefore we are interested in the conductivity along the local field gradient. But the linearity (one-dimensionality) of the current in some piece of metal does not contradict the three-dimensionality of the volume itself, filled with metal. For example, in [19] the calculation was carried out in the approximation (in fact, one-dimensional) of a thin wire. Thus, a significant local anisotropy of the electrical conductivity makes it possible to use the model of the quasi-one-dimensional conduction band at the microscopic level for approximate calculations of this parameter for a local region.

Since the tunneling transparency coefficient has a nonlinear character, the average value of such a function differs markedly from the non-fluctuating value [33]. The conductivity has a resonant-percolation nature: various short sections with metallic conductivity arise by maintaining a short-range order. Despite the movement of chaotically located particles relative to each other, the Debye-Waller factor is negligible (exponentially small) due to the relative smallness of wind velocities compared to speeds of lightnings. In a purely one-dimensional case, Anderson's localization would take place, but in reality we have a three-dimensional case. Lightning passes in such areas (along the path of least resistance), sometimes punching their way between the nearest segments. As an example of such approach, some suprabarrier propagation condition for the outer electron of a negative ion is presented in [32]:

$$\frac{4\pi}{3}\left(\frac{\alpha_a e^2}{E_i}\right)^{3/4} n_a \approx \frac{1}{3},$$

where $E_i$ - is the energy of the electron affinity of an atom, $\alpha_a$ - is polarizability of an atom, $n_a$ - is the concentration of atoms.

To describe properties of the metallic bond and formation of the metal channel in our case, it is necessary to consider that external excess charges create this bond, and it is required to take into account their interaction with the surrounding matter. Due to the relatively low concentration of excess charges, the one-particle approximation is applicable. Data on electron affinities vary considerably in physical and chemical literature; therefore chosen values have evaluative character. So, for an isolated molecule of water we can take the energy value of the electron affinity -0.9 eV (there is no stable ion in this case, but charged clusters are formed). For an isolated oxygen molecule, a value of -0.87 eV can be chosen (in most handbooks an even smaller value is given: -0.44 eV). For a nitrogen molecule, this energy is close to zero. Describing the Catatumbo Lightning, it must be taken into account the additional specific gases released in a given place with low-energy electron affinity (for example, methane with zero affinity energy), which "dilute" other molecules.

Thus, the magnitude of barriers at capture of an additional electron does not exceed 0.9 eV in the case of neutral particles (since percolation does not have to pass through the top of the barrier): $U_0 \leq 0.9$ eV. The width of a single barrier is of the order of the molecular diameter: $d \leq 3.2$ nm. Depending on the humidity, temperature and turbulence, the average number of oxygen atoms in a linear chain can exceed the number of water vapor molecules in the tens and hundreds of times. In addition, it is necessary to take into account the presence of incorporated ions (in the area of formation of lightnings - mostly negative) at some averaged distances from each other. If the charge of certain crystals is $q_- \approx -4 \cdot 10^{-13}$ C, their contribution to the potential energy of the selected charge become $\sim +1$ eV, i.e. the final retention potential becomes zero. Thus, it will be closer to the lower limit $0 \leq U_0 \leq 0.9$ eV on the average along the resonance trajectory. Note that as a first approximation the Wulff - Bragg condition $2d \sin \theta = n\lambda$, rewritten as $2d \sin \theta = 2\pi n v \hbar / E$, allows take into account the change of the energy of conduction bands for the non-rectilinear propagation of electrons (for bayable at some angle sections of resonant trajectories). For discharge propagation, it is sufficiently to join conduction bands of neighboring resonance sections accurate to the energy of thermal fluctuations (or electrical noise). In the presence of such different choices, the number of possibilities for the formation of a metal bond (resonant paths) is sufficiently large.

Since it is impossible to describe such a complex nonequilibrium system rigorously, we illustrate some of the above-mentioned moments by considering two limiting cases: 1) resonance percolation over a small number of barriers, which can also eventually "form" into a fractal grid, and 2) temporarily arising ordered substructures. For example, the transparency of the system of two identical barriers located at a certain distance can be equal to one for a certain energy $E$. Such a resonant length $L$ for the rectangular barriers $U$ of the width $a$ are determined by the formula:

$$L = \frac{2\pi n}{1{,}0798519\sqrt{E}} \pm \frac{\arccos\left[\dfrac{U^2 - (8E^2 - 8EU + U^2)\cosh(1{,}0798519 a\sqrt{U-E})}{8E^2 - 8EU + U^2 - U^2 \cosh(1{,}0798519 a\sqrt{U-E})}\right]}{1{,}0798519\sqrt{E}}.$$

Here the energy is expressed in electron volts, and the distance - in angstroms; it is necessary to take the upper sign (+) and $n \geq 0$ for $E > U/2$, but the lower sign (-) and $n \geq 1$ is taken when $E < U/2$ (You can check in the symbolic program that the substitution of this value identically leads the transparency $D$ of the system of barriers to one).

We give numerical estimates for molecules. For a molecule of water: $U = 0.9$ eV, $a = 3$ Å. If the energy of an excess electron is 0.3 eV, the resonance distance between water molecules is $L_1 = 6.3$ Å. For an electron with the energy of 0.09 eV the resonance distance is equal to 15.097 Å. Similarly, we take for oxygen: $U = 0.87$ eV, $a = 3.2$ Å. For $E = 0.3$ eV the resonance distance $L_2$ between the molecules of oxygen equals 6.24 Å. And for an excess electron with an energy of 0.09 eV the resonance distance is 15.067 Å. We now observe that for any arbitrary distance $L$ from such a "resonant pair" to another "resonant pair" (similar to or composed of other molecules), the given energy will remain resonant energy for the summary disordered system. Further, electrons are not obliged to move along highs of barriers (because electrical current tries to choose the easiest way). Besides, even with all the fixed distances, there exist other resonance energies (in addition to the arbitrary chosen value). This is illustrated in Figure 1. It shows the total transparency $D$ of a system consisting of two different "resonant pairs" (0.3 eV arbitrarily chosen as the resonant energy), depending on the energy $E_0$ of an excess electron and on the distance $L$ between the pairs. We also note that the "barrier" ideology is successfully applied not only in quantum mechanics, but also in the study of the passage of wave barriers in such a heterogeneous (and nonstationary) medium, like plasma.

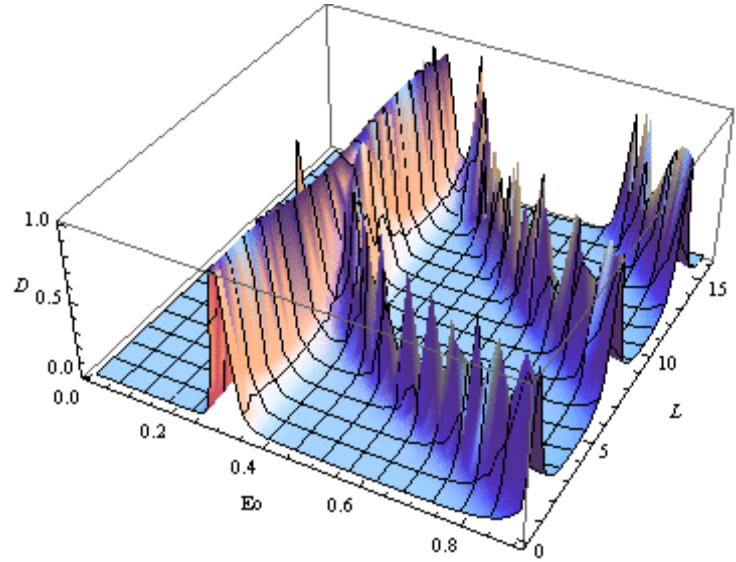

**Fig. 1** The transparency of two pairs of barriers

Thus, in principle, the formation of metal bonds in a thundercloud can be explained by percolation even in a system with disorder (there are numerous examples of so-called colored processes, when the intense, narrow separated lines are observed in the spectra of disordered media and complex chaotic processes). We believe that the short-range order can randomly (dynamically) arise from the neighboring barriers that are at a resonant distance (pair of barriers). Since these pairs can provide resonant passage of particles, they are likely to be more resistant to external influences.

Now we will analyze another possibility (the second case): the temporary formation of local quasiperiodic substructures. For a quantitative description of the collective properties of parts from which will be drawn up the metal-like structure of storm clouds, we use the generally accepted one-electron approximation from the theory of metals, that is, we consider the Schrodinger equation

$$\hat{H}\psi_{\mathbf{k}}(\mathbf{r}) = E\psi_{\mathbf{k}}(\mathbf{r})$$

for an electron moving in the self-consistent field (quasi-periodic potential). It is known that solutions of this equation are the Bloch functions:

$$\psi_{\mathbf{k}}(\mathbf{r}) = u_{\mathbf{k}}(\mathbf{r})\exp(i\mathbf{k}\mathbf{r}),$$

where $u_{\mathbf{k}}(\mathbf{r})$ - is a periodic function with the period of "a statistical lattice", $\mathbf{k}$ - quasiwave vector. Let us recall that an electron is characterized by a quasi-momentum $\mathbf{P} = \hbar \mathbf{k}$ and an energy $E(\mathbf{k}) = E(\mathbf{k} + 2\pi \mathbf{H})$, where $\mathbf{H}$ - is the vector of the reciprocal "lattice". To completely describe the totality of electron states in this "accidentally arising section of the metal crystal", it is enough to consider the range of values of $\mathbf{k}$ bounded by the first Brillouin zone only.

Methods for calculating the Bloch functions are well known [34, 35]. We will solve the problem numerically in the one-dimensional case (in fact, along the resonant-percolation path). For the numerical solution we use the algorithm and program in Mathematica from [36]. There the formula was used

$$\cos(ka) = \frac{\cos(\theta + \chi a)}{|t|}, \qquad (1)$$

where $a$ - is the period of "a lattice", the amplitude reflection coefficient is $t = |t|e^{i\theta}$ (this coefficient is defined as in the scattering problem), the energy of an electron is

$$E = \frac{\hbar^2 \chi^2}{2m}. \qquad (2)$$

On the discrete set of points $x_n = n\Delta$ ($n=0, 1, ..., N$) for a wave function $\psi_n = \psi(x_n)$, it have searched for the recursion solution

$$\psi_{n+1} = R_n \psi_n$$

of the Schrodinger equation $\psi_{n+1} + \psi_{n-1} + u_n \psi_n = 0$, where

$$u_n = -2 + \frac{2m\Delta^2}{\hbar^2}(E - V_n), \qquad (3)$$

$V_n = V(x_n)$ - is a potential. It turned a recurrence relation

$$R_{n-1} = -\frac{1}{u_n + R_n} \qquad (4)$$

with the boundary condition

$$R_{N-1} = -\frac{1}{u_N/2 + i\chi\Delta}. \tag{5}$$

The initial condition

$$\psi_0 = \frac{2i\chi\Delta e^{-i\chi a/2}}{R_0 + (u_0/2 + i\chi\Delta)} \tag{6}$$

allows us to find all functions in nodes. Then the amplitude reflection coefficient is:

$$t = \psi_N e^{-i\chi a/2}. \tag{7}$$

The algorithm for calculating is the following. Choosing the potential $V(x)$ and a tunnelling electron energy $E$, we calculate $\chi, u_n, V_n, R_{N-1}$. Then we solve the equation for the band spectrum (1) under condition

$$\left|\frac{\cos(\theta + \chi a)}{|t|}\right| \leq 1.$$

The potentials we approximated by dependencies:

$$U_i = \frac{U_{0i}(r - r_i)^2}{c_i^2 + (r - r_i)^2}\delta,$$

with a limiting factor $\delta = (1 - \text{sign}[r - r_i - a/2]\text{sign}[r - r_i + a/2])/2$. Here $U_{0i}$ - is the value of the potential of i-th molecule, $r_i$ - is its coordinate in the elementary cell, $c_i$ determines the width of the potential well (we use $0.1a\sqrt{U_{0i}}$).

Let us give an example of numerical calculations (distances are taken in Angstroms, energy - in electron volts). Figure 2 qualitatively illustrates the energy bands for droplet of water, Figure 3 corresponds to the ice crystal, Figure 4 shows the energy zones of the air. Since it comes exclusively about the excess electrons (their number is much smaller than the number of molecules in thunderstorm clouds), even the first zone will be almost empty: the Fermi energy

$$E_F(0) \approx \frac{\hbar^2}{2m}(3\pi^2 N)^{2/3} \approx 10^{-12} \text{ eV}.$$

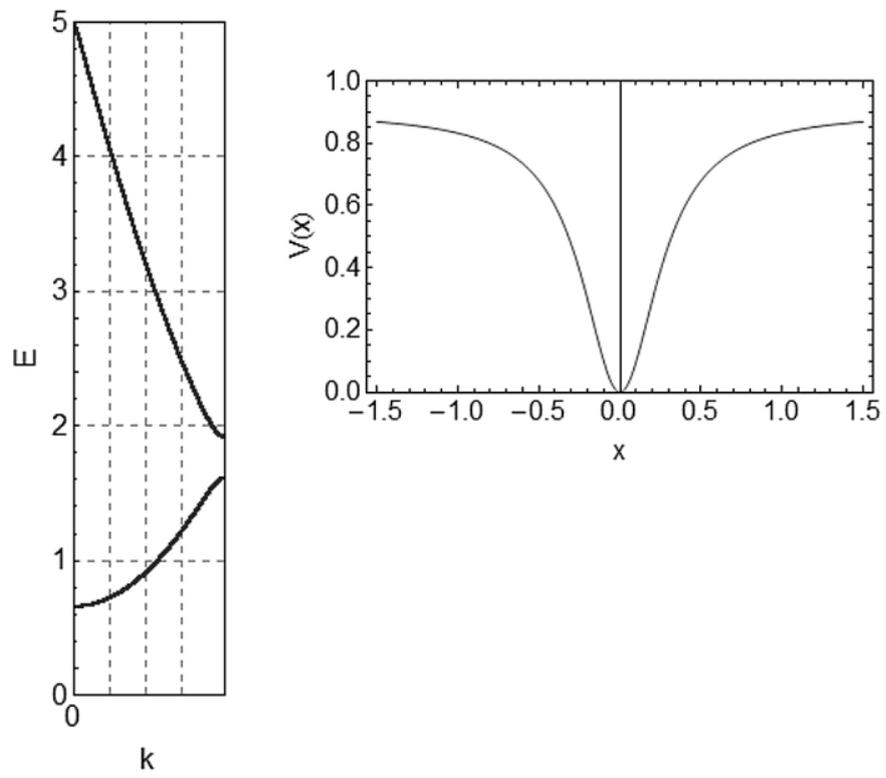

**Fig. 2** The potential (the electron affinity) and the energy bands of a water drop for the excess electrons

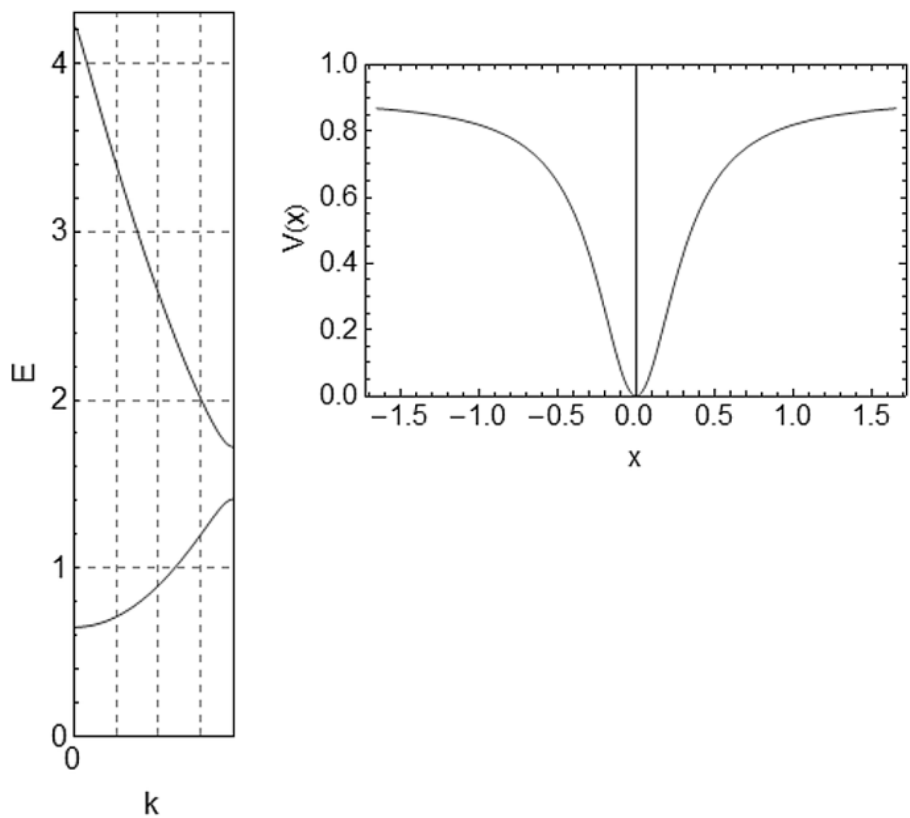

**Fig. 3** The potential (the electron affinity) and the energy bands of ice crystals for the excess electrons

Energy bands of air, water and ice droplets are different and, in their docking, create ample opportunities for sequential increase of the electron energy in the field, starting with a certain threshold, which is less than the breakdown value.

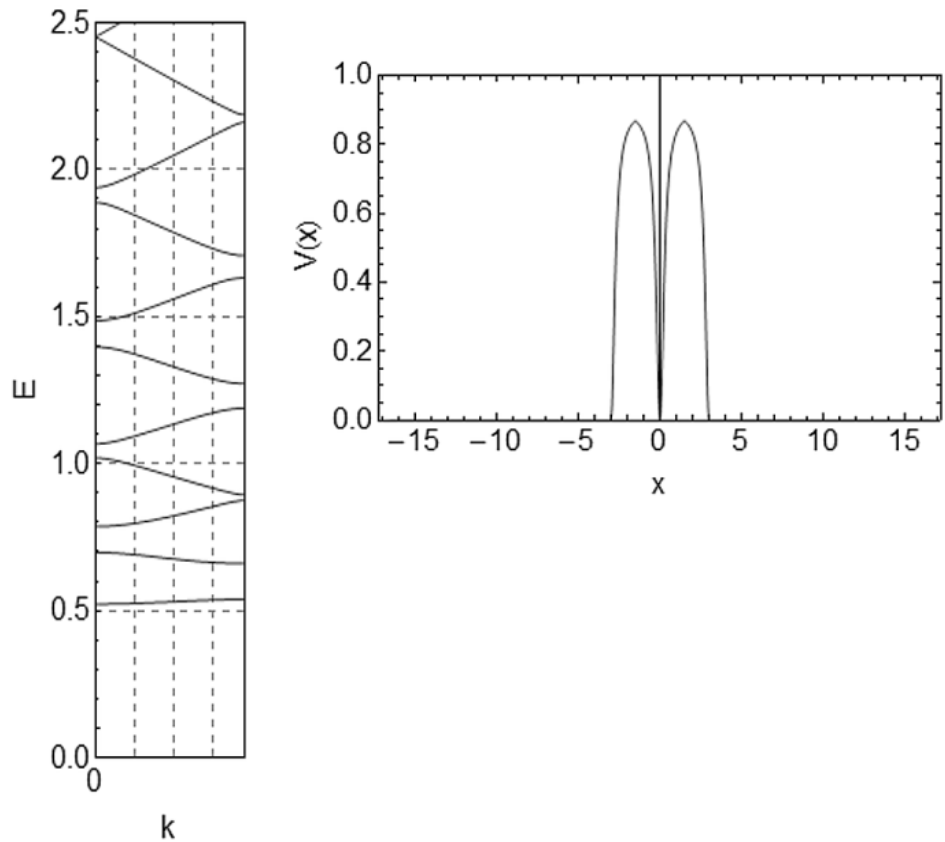

**Fig. 4** The potential (the electron affinity) and the energy bands of oxygen for the excess electrons

Let us remind that in a field of 250 kV/m an electron over a length of its path can gain energy ~0.3 eV, which is enough to get into the conduction band. Numerical estimates show that in this conduction band the effective mass of an electron is

$$m^* = \hbar^2 \left[ \frac{d^2 E}{dk^2} \right]^{-1} \approx 1,046 m_e$$

(for the lower band this value is approximately 3-4 times greater).

As an example, Figure 5 provides one of many possible final pictures with a real potential, taking into account an excess charge in thunderclouds (this experimental value of the order of $10^{-8}$ C/m$^3$). Of course, the picture is only illustrative in nature (because there are a huge number of other favorable opportunities within the real cloud characteristics). Once again, we emphasize that no infinite ordering exists in nature at all (this is a mathematical way of describing in the band theory only). But for practice, we do not need a continuous zone. It is sufficient that the distances between the nearest energy levels inside the zone become of the order of the thermal or electric fluctuation energy, then an electron can wander between them (this will be the case of the band no longer for an infinite periodic structure, but for a finite quasiperiodic substructure). It should be noted that there is nothing unusual in the application of such approaches: the ideology of the mean field is rather fruitful in quantum mechanics. In addition, even in the theory of turbulence, the Fourier expansion of wave packets leads to a regular lattice [37]. As some confirmation, we can also cite work on plasma crystals [38-42].

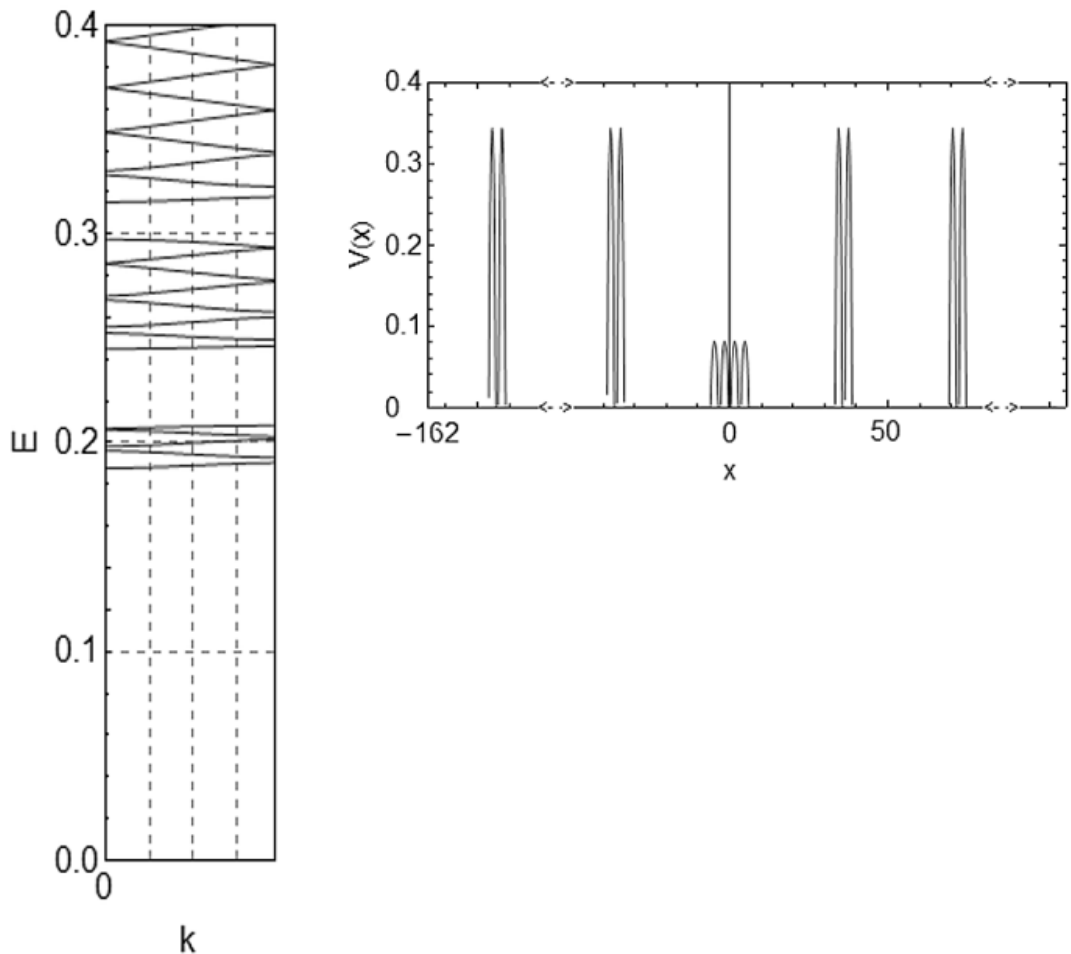

**Fig. 5** The potential and an approximation of the energy bands in a thunderstorm cloud for the excess electrons

Balloon and aircraft measurements record the presence of charged layers inside thunderstorm clouds and average electric field strengths up to hundreds of kV/m. Remote measurements of radio emission show that, due to turbulent motions, the electric field of a thunderstorm cloud is spatially inhomogeneous and fluctuates [43]. Then qualitatively the lightning discharge mechanism is the following. During the streaming separation of charges, the electric field strength is gradually increasing between reinforcing charged regions (where the number of the charges of the same sign on crystals are increasing). As a result, an increasing number of ice crystals in the local areas between charged cloud subsystems receives a preferential orientation (statistically) in the direction of the local field. A locally ordered structure from directed crystals and ions arises. Since the negatively charged crystals (whose charge is concentrated on the surface) have low retention potential, after the turn along the field they as though drastically "dilute" the shortest path of electrons by more transparent portion. Linear paths along crystals become resonant upon reaching a certain concentration of charged particles and of field intensity (less than the breakdown value). There arises instability. It is known that the presence of sharp forms can increase the efficiency of the electron emission by an order of magnitude. Excess electrons "slide" along planes of crystals toward the oppositely charged region on the arisen (resonant-percolation) metal area. Part of the charge through "a metal grid" of the charged cloud mass flows into this arising channel. Herewith the local field strength on the front portion of lightning is higher than the mean value that supports the instability and allows carrying the breakdown between the nearest metallic regions. Of course, the EAS and the runaway breakdown can contribute to the ionization and initiation of the process.

The ratio of several factors is important for the process of orientation of crystals. The Brownian motion randomizes crystals. Aerodynamic lifting force counteracts the force of gravity, supporting the crystals in the air at a certain height. The moment of aerodynamic lifting force is trying to orient the crystals parallel to the ground (by larger surface) with $\theta_2 = \pi/2$, where $\theta_2$ - is the angle between the plane of the largest cross-section and the flow direction **V**. The electric force participates in the establishment of equilibrium motion. The moment of electric force tends to turn the crystals along the field **E** with $\theta_1 = 0$, where $\theta_1$ - is the angle between the field direction and the greatest length of the crystal. As a result, the statistical probability to have the orientation of the axis of symmetry for the crystal under a certain angle to the field is given by:

$$P(\theta_1)d\theta_1 = b\exp\{-[U_{el}(0)\cos^2\theta_1 + U_a(0)\sin^2\theta_2]/(kT)\}d\theta_1,$$

where $b$ – is the normalization constant, $U_{el}(0) = f_1 E^2$, $U_a(0) = f_2 V^2$, $f_1$ and $f_2$ - depends on the geometric characteristics of the crystal.

Taking into account the variability of thunderstorm fields, the change of the orientation of crystals in the function of time is determined by the equation:

$$\frac{\partial P}{\partial t} = \frac{A}{\sin\theta}\frac{\partial}{\partial\theta}\left[\sin\theta\left(\frac{\partial P}{\partial\theta} + \frac{\mu E_x(t)\sin\theta}{kT}P\right)\right] + \frac{A}{\sin^2\theta}\frac{\partial^2 P}{\partial\varphi^2},$$

where $A = f_3 kT/(\eta a^3)$, $\eta$ - is the viscosity of the medium, $f_3$ - is a factor depending on the shape of crystals, $a$ - is their characteristic size, and $\mu$ - is momentum.

Calculations and experiments [44, 45] show that the probability of ordering decreases with decreasing crystal size. Small crystals (up to several micrometers) are randomized by the Brownian motion, but medium and large ones are ordered in the presence of field of $50-100$ kV/m. The ordering occurs in a dozen microseconds that initiates the formation of a resonant channel, while the orientation of the crystals tracks the rapid changes in the field at each local site. The minimum length of lightning is defined as follows. Only far enough separated (by aerodynamic forces), charged cloud subsystems can be strong enough (with the highest total charge and its concentration) that to begin the formation of metal bonds within the subsystems and to begin the above described process of formation of the lightning channel (the initiation of lightning). Thus, the field of 250 kV/m is sufficient for the initiation of lightning discharges. If there is such the gas composition of atmosphere in a storm that the energy spectra are joined in a continuous manner, then the energy of electrons during lightning flash is growing continuously without breakdowns of forbidden zones (without thunder - as this is in the Catatumbo Lightning).

The average distance between particles in thunderstorm clouds is of the order of 1 mm. For particles with a size of ~ $20-80$ μm, the average change in the distance for ordering by the field is $d = (l_{max} - l_{min})/2 \approx 0.025$ mm. Earlier we saw that the transparency of the barrier system increases sharply to unity for small changes (~ 0.1*L*) near the resonance distance *L*. In the first approximation, we assume that the conductivity is proportional to the product of the transparency coefficients of the barriers $\sigma \propto \prod_{i=1}^{N} D_i$, where $N \approx 10^3$. Since the resonance distances are much smaller than *d*, the relative increase in conductivity $\sigma/\sigma_0 = 1/D_0^N$ could reach tens of orders (as an example, the translucence of the potential barrier and the formation of a correlated coherent state for the model potential are considered more rigorously in [46, 47]). In fact, at voltages lower than the breakdown values, a metal channel appears, a lightning strike occurs. Partial discharge of charge leads to voltage drop and partial disordering of the crystals. Further the process can be repeated many times.

Note the following. The more ions are in the atmosphere and the larger is the average distance between molecules, the smaller is the resonance energy (levels of zones are decreased). This leads to a significant reduction in the minimum field strength required for the discharge. Increasing the number of ions also contributes to a more rapid equalization of potential on the surface perpendicular to the discharge. Perhaps, abovementioned circumstances play an important role for high-altitude lightnings occurring for the same reason is not by narrow channels, but by relatively wide flows. However, this problem requires separate consideration, and this article is not about such phenomena.

Summing up what has been said, we emphasize that the article does not state that the cloud substance is converted into metal, as if the whole thunder cloud is solidly connected by a metallic bond, and the resulting metal bond is itself more stable and exists for a long time. In the article, the viewpoint is argued that in a thundercloud (in the presence of an excess charge and a strong electric field), conditions arise for repeatedly repeating local manifestations of the metallic bond. Such the bond, each time suddenly appearing just before the lightning discharge, activates and "organizes" the next lightning channel and some chaotic conductive network connecting this channel with the nearest cloud space. With the end of the lightning discharge, the bond is partially or completely destroyed. Otherwise, there would be a uniform neutralization of the cloud. For this phenomenon, metallization has nothing to do with the electrons of the outer shells of atoms. The excess electrons that are retained due to the affinity energy to the electron are collectivized. These collectivized electrons are capable of increasing their energy in an external electric field that, from the viewpoint of quantum mechanics, is a sign of the presence of a metallic bond.

## 4 Conclusion

The paper discussed some observations concerning the electrical phenomena in the atmosphere. Some statistical characteristics of the atmospheric electrical activity were analyzed, and it was concluded that the Earth magnetic field impacts on them. On the basis of experimental data on the thunderstorm activity, a mechanism was considered which allows charges from the large charged cloud subsystems to concentrate in the lightning channel during the implementation of lightning flashes. The short-range order can randomly (dynamically) arises from neighboring barriers that are at a resonant distance (barrier pairs). These can be charges (ions) frozen into one and the same or different crystals that have appeared at a resonant distance. Since such pairs can provide resonant passage of particles, they are likely to be more resistant to external influences. We assume that such pairs of barriers can support the resonance-percolation nature of the conductivity when the static electricity concentration is within certain limits (and the characteristics of cloud particles are also within certain limits). Some distance correlation exists only within the pair, and relative to each other these pairs can be located and move chaotically. When a certain concentration of such pairs is reached, a chaotic resonant grid arises and the possibility for the charges to flow to the lightning channel is realized. There can also be a significant increase in the number of resonance paths and metallization of the channel. The formation of metallic bonds in thunderclouds is described as the second-order phase transition. The new mechanism for the formation and distribution of the lightning channel was suggested. This mechanism is associated with the successive occurrence of portions with metal bonds (by the resonant-percolation manner) as a result of the sudden increase in the linear charge concentration and changes in characteristics of the trajectory, caused by a turn of charged crystals along the field (by ordering) in increasing electric field. The local increase of the field, in its turn, supports the formation and further propagation of the metallic channel (in which propagates the lightning).

**References**

1. Wilson, C.T.: Investigations on Lightning Discharges and on the Electric Field of Thunderstorms. *Philos. Trans. Roy. Soc. Lond. A*. **221**, 73-115 (1920).

2. Mareev, E.A.: Global electric circuit research: achievements and prospects. *Phys. Usp*. **53**, 504-511 (2010). doi:10.3367/UFNe.0180.201005h.0527

3. Harrison, R.G.: The Carnegie Curve. *Surv. Geophys*. **34**, 209-232 (2013).

4. Muchnik, V.M.: Fizika grozy (Physics of thunderstorm). Gidrometeoizdat, Leningrad (1974).

5. Elster, J., Geitel, H.: Zur Influenztheorie der Nieder- schlagselektrizität. *Phys. Z.* **14**, 1287-1292 (1913).

6. Wilson, C.T.R.: Some thundercloud problems. *J. Franklin Inst*. **208**, 1-12 (1929).

7. Grenet, G.: Essai d'explication de la charge electrique des nuages d'orages. *Ann. Geophys*. **3**, 306-307 (1947).

8. Vonnegut, B.: Possible mechanism for the formation of thunderstorm electricity. Geophys. Res. Paper No. 42. Proc. Conf. Atmos. Electr. AFCRC-TR-55-222, 169-181 (1955).

9. Chalmers, J.A.: Atmospheric Electricity. Pergamon Press, New York (1967).

10. Avila, E.E., Longo, G.S., Burgesser, R.E.: Mechanism for electric charge separation by ejection of charged particles from an ice particle growing by riming. *Atmos. Res.* **69**, 99-108 (2003).

11. Mason, J., Mason, N.: The physics of a thunderstorm. *Eur. J. Phys.* **24**, S99-S110 (2003).

12. Handel, P.H.: Polarization catastrophe theory of cloud electricity - speculation of a new mechanism for thunderstorm electrification. *J. Geoph. Res.: Atmos.* **90** D3, 5857–5863 (1985).

13. Bazelyan, E.M., Raizer Ju.P.: Fizika molnii I molniezaschity (Physics of lightning and lightning protection). Fizmatlit, Moscow (2001).

14. Rakov, V.A., Uman, M.A.: Lightning: Physics and Effects. Cambridge University Press, New York (2003).

15. Iudin, D.I., Trakhtengerts, V.Yu.: Fractal Dynamics of Electric Charges in a Thunderstorm Cloud. Izvestiya, *Atmospheric and Oceanic Physics*. **36**, 597-608 (2000).

16. Iudin, D.I., Trakhtengertz, V.Yu., Hayakawa, M.: Fractal dynamics of electric discharges in a thundercloud. *Phys. Rev. E*. **68**, 016601 (2003).

17. Iudin, D.I., Trakhtengertz, V.Yu.: Sprites, Elves and Intense Lightning Discharges. *NATO Science Series. Springer*. **225**, 341-376 (2006).

18. Lowke, J.J.: The initiation of lightning in thunderclouds: The possible influence of metastable nitrogen and oxygen molecules in initiating lightning streamers. *J. Geophys. Res. Atmos*. **120**, 3183-3190 (2015).

19. Carlson, B.E., Liang, C., Bitzer, P., Christian, H.: Time domain simulations of preliminary breakdown pulses in natural lightning. *J. Geophys. Res. Atmos*. **120**, 5316-5333 (2015).

20. Artekha, S.N., Belyan, A.V.: On the role of electromagnetic phenomena in some atmospheric processes. *Nonlin. Processes in Geophys*. **20**, 293-304 (2013). doi: doi.org/10.5194/npg-20-293-2013

21. Artekha, S.N., Belyan, A.V., Erokhin, N.S.: Manifestations of electromagnetic phenomena in atmospheric processes. *Sovr. Probl. DZZ Kosm.* **10(2)**, 225-233 (2013).

22. Arteha, S.N., Belyan, A.V., Erokhin, N.S.: Electromagnetic phenomena in atmospheric plasma-like subsystems. *Problems of Atomic Science and Technology*. **4(86)**, 115-120 (2013).

23. Marshall, T.C., Stolzenburg, M., Maggio, C.R., Coleman, L.M., Krehbiel, P.R., Hamlin, T., Thomas, R. J., Rison, W.: Observed electric fields associated with lightning initiation. *Geophys. Res. Lett*. **32**, L03813 (2005).

24. Gurevich, A.V., Milikh, G.M., Roussel-Dupre, R.: Runaway electron mechanism of air breakdown and preconditioning during a thunderstorm. *Physics Letters A*. **165(5-6)**, 463-468 (1992).

25. Gurevich, A.V., Zybin, K.P.: Runaway breakdown and electric discharges in thunderstorms. *Phys. Usp*. **44** 1119–1140 (2001).

26. Gurevich, A.V., Karashtin, A.N., Ryabov, V.A., Chubenko, A.P., Shchepetov, S.V.: Nonlinear phenomena in the ionospheric plasma. Effects of cosmic rays and runaway breakdown on thunderstorm discharges. *Phys. Usp*. **52**, 735–745 (2009).

27. Lifshitz, I.M., Kirpichenkov, V.Ya.: Tunnel transparency of disordered systems. *JETP*, **50(3)**, 499-511 (1979).

28. Lifshits, I.M., Gredeskul, S.A., Pastur, L.A.: Introduction to the theory of disordered systems. Wiley, New York (1988).

29. Artekha, S.N., Belyan, A.V.: On a possible mechanism for lightning flashes. Proceedings of the International Conference MSS-14 "Mode Conversion, Coherent Structures and Turbulence". 58-63, LENAND, Moscow (2014).

30. Zaitsev, R.O., Kuz'min, E.V., Ovchinnikov, S.G.: Fundamental ideas on metal-dielectric transitions in 3d-metal compounds. *Physics - Uspekhi* **29**, 322–342 (1986).

31. Hartree, D.R.: The wave mechanics of an atom with a non-coulomb central field. Part I. theory and methods. *Mathematical Proceedings of the Cambridge Philosophical Society*. **24**, 89-110 (1928). doi:doi.org/10.1017/S0305004100011919

32. Likal'ter, A.A.: Gaseous metals. *Physics - Uspekhi* **35(7)**, 591–605 (1992).

33. Artekha, S.N., Moiseev, S.S.: Transmission of randomly nonuniform barriers and certain physical consequences. *Technical Phys*. **38**, 265-271 (1993).

34. Kittel, Ch.: Introduction to solid state physics. Wiley, New York (1971).

35. Ashcroft, N.Q.W., Mermin, N.D.: Solid State Physics. Saunders College, Philadelphia (1976).

36. Denisenko, M.V., Derbenko, A.S., Kashin, S.M., Satanin, A.M.: Calculation of the Bloch functions of an electron in a one-dimensional periodic potential. NNSU, Nizhny Novgorod (2010).

37. Nakano, T.: Direct interaction approximation of turbulence in the wave packet representation. *Phys. Fluids*. **31(6)**, 1420-1430 (1988).

38. Thomas H., Morfill G.E., Demmel V., Goree J., Feuerbacher B., Möhlmann D.: Plasma crystal: Coulomb crystallization in a dusty plasma. *Phys. Rev. Lett*. **73(5)**, 652-655 (1994).

39. Chu, J.H, Lin, I.: Coulomb lattice in a weakly ionized colloidal plasma. *Physica A*. **205(1-3)**, 183-190 (1994).

40. Tsytovich, V.N.: Dust plasma crystals, drops, and clouds. *Phys. Usp*. **40**, 53-94 (1997). doi:10.1070/PU1997v040n01ABEH000201

41. Fortov V.E., Khrapak A.G., Khrapak S.A., Molotkov V.I., Petrov O.F.: Dusty plasmas. *Phys. Usp*. **47**, 447-492 (2004). doi:10.1070/PU2004v047n05ABEH001689

42. Morfill, G.E., Ivlev, A.V.: Complex plasmas: An interdisciplinary research field. *Rev. Mod. Phys*. **81**, 1353-1404 (2009). doi:doi.org/10.1103/RevModPhys.81.1353

43. Iudin, D.I., Iudin, F.D., Hayakawa, M.: Modeling of the intracloud lightning discharge radio emission. *Radiophysics and Quantum Electronics*. **58(3)**, 173–184 (2015).

44. Saunders, C.P.R., Rimmer, J.S.: The electric field alignment of ice crystals in thunderstorms. *Atmos. Res.* **51**, 337-343 (1999).

45. Foster, T.C., Hallet, J.: The alignment of ice crystals in changing electric fields. *Atmos. Res.* **62**, 149-169 (2002).

46. Vysotskii, V.I., Vysotskyy, M.V., Adamenko, S.V.: Formation and application of correlated states in nonstationary systems at low energies of interacting particles. *Journal of Experimental and Theoretical Physics*. **114(2)**, 243–252 (2012).

47. Vysotskii, V.I., Vysotskyy, M.V.: Correlated states and transparency of a barrier for low-energy particles at monotonic deformation of a potential well with dissipation and a stochastic force. *Journal of Experimental and Theoretical Physics*. **118(4)**, 534-549 (2014).